\documentclass[reprint,aps,prd,longbibliography,floatfix,superscriptaddress,showkeys]{revtex4-2} 
\usepackage{amsmath}
\usepackage{amsfonts}
\usepackage{amssymb}
\usepackage[hidelinks]{hyperref}
\usepackage{tabulary}
\usepackage[dvipsnames]{xcolor}
\usepackage{soul}
\usepackage{siunitx} %
\usepackage{selinput}%
\usepackage{todonotes}
\usepackage{cleveref} %
\usepackage{upgreek} %

\usepackage{tikz}
\usepackage{graphicx}
\usepackage{placeins} %
\usepackage{blindtext}
\usepackage{tabulary}
\usepackage{rotating}
\usepackage{afterpage}

\usepackage{multirow}
\usepackage{makecell}

\usepackage{booktabs}

\usepackage[toc,xindy,nonumberlist,shortcuts,acronym,automake]{glossaries}
\usepackage{csquotes}

\newacronym{AOM}{AOM}{acousto-optic modulator}
\newacronym{ASD}{ASD}{amplitude spectral density}

\newacronym{BH}{BH}{black hole}

\newacronym{CoM}{CoM}{centre of mass}

\newacronym{DF}{DF}{drag-free}
\newacronym{DFACS}{DFACS}{drag-free attitude control system}
\newacronym[firstplural={degrees of freedom (DOFs)}]{DOF}{DOF}{degree of freedom}
\newacronym{DMU}{DMU}{data management unit}
\newacronym{DPLL}{DPLL}{digital phase locked loop}
\newacronym{DRS}{DRS}{disturbance reduction system}
\newacronym{DWS}{DWS}{differential wavefront sensing}

\newacronym{EH}{EH}{electrode housing}
\newacronym{EOM}{EOM}{electro-optical modulator}
\newacronym{ESA}{ESA}{European Space Agency}

\newacronym{FIOS}{FIOS}{fibre injector optical subassembly}

\newacronym{GRACE-FO}{GRACE-FO}{Gravity Recovery and Climate Experiment Follow On}
\newacronym{GRS}{GRS}{Gravitational Reference Sensor}
\newacronym{GW}{GW}{gravitational wave}

\newacronym{IFO}{IFO}{interferometer}
\newacronym{IMS}{IMS}{interferometric measurement system}

\newacronym{LISA}{LISA}{Laser Interferometer Space Antenna}
\newacronym{LPF}{LPF}{LISA Pathfinder}
\newacronym{LPS}{LPS}{longitudinal path length signal}
\newacronym{LTPDA}{LTPDA}{LPF Data Analysis package}
\newacronym{LTP}{LTP}{LISA Pathfinder technology package}
\newacronym{LXE}{LXE}{long cross-talk experiment}

\newacronym{MCMC}{MCMC}{Markov-chain Monte Carlo}

\newacronym[firstplural={Optical Benches
  (OBs)}]{OB}{OB}{Optical Bench}
\newacronym{OMS}{OMS}{Optical Metrology System}

\newacronym{PD}{PD}{photodiode}
\newacronym{PDS}{PDS}{photo detection system}
\newacronym{PLL}{PLL}{phase-locked loop}
\newacronym{PM}{PM}{phasemeter}
\newacronym{PSD}{PSD}{power spectral density}

\newacronym{QPD}{QPD}{quadrant photodetector}

\newacronym{REF}{REF}{reference}
\newacronym{RIN}{RIN}{relative intensity noise}

\newacronym[firstplural={spacecraft
  (S/C)},plural=S/C]{S/C}{S/C}{spacecraft}
\newacronym{SEPD}{SEPD}{single-element photo diode}
\newacronym{SNR}{SNR}{signal-to-noise ratio}

\newacronym{TBC}{TBC}{to be confirmed}
\newacronym{TBD}{TBD}{to be determined}
\newacronym{TDI}{TDI}{time-delay interferometry}
\newacronym[first={test mass (TM)}]{TM}{TM}{test mass}
\newacronym{TS}{TS}{telescope}
\newacronym{TTL}{TTL}{tilt-to-length}

\newacronym{UTC}{UTC}{coordinated universal time}

\makeglossaries

\usepackage{ulem}
\usepackage{todonotes}

\begin{document}
\title{Tilt-to-length coupling in LISA Pathfinder: analytical modelling}
\author{Marie-Sophie Hartig}
\email{marie-sophie.hartig@aei.mpg.de}
\affiliation{Max Planck Institute for Gravitational Physics (Albert Einstein Institute)}
\affiliation{Institute for Gravitational Physics of the Leibniz Universit\"at Hannover} 
\author{Gudrun Wanner}
\email{gudrun.wanner@aei.mpg.de}
\affiliation{Institute for Gravitational Physics of the Leibniz Universit\"at Hannover} 
\affiliation{Max Planck Institute for Gravitational Physics (Albert Einstein Institute)}

\begin{abstract}
Tilt-to-length coupling was the limiting noise source in LISA Pathfinder between 20 and 200\,mHz before subtraction in post-processing. 
To prevent the adding of sensing noise to the data by the subtraction process, the success of this strategy depended on a previous direct noise reduction by test mass alignment.
The exact dependency of the level of tilt-to-length coupling on the set-points of LISA Pathfinder's test masses was not understood until the end of the mission.
Here, we present, for the first time, an analytical tilt-to-length coupling model that describes the coupling noise changes due to the realignments. 
We report on the different mechanisms, namely the lever arm and piston effect as well as the coupling due to transmissive components, and how they contribute to the full coupling. 
Further, we show that a pure geometric model would not have been sufficient to describe the coupling in LISA Pathfinder. Therefore, we model also the non-geometric tilt-to-length noise contributions.
For the resulting coupling coefficients of the full model, we compute the expected error bars based on the known individual error sources. 
Also, we validated the analytical model against numerical simulations.
A detailed study and thorough understanding of this noise are the basis for a successful analysis of the LISA Pathfinder data with respect to tilt-to-length coupling.
\end{abstract}
\maketitle

\section{Introduction}

\begin{figure}
  \includegraphics[width=\columnwidth]{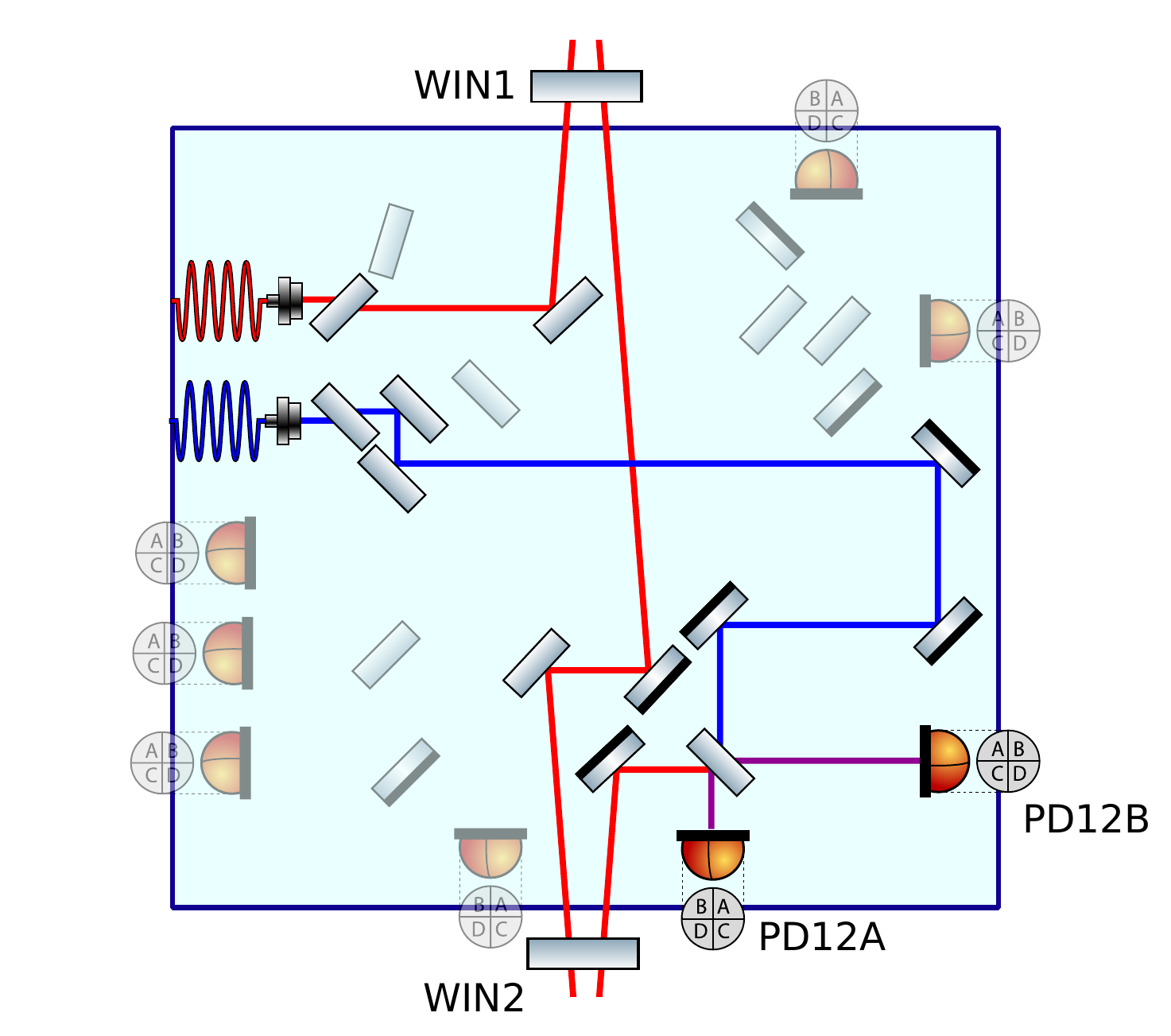}
  \caption{The x12-interferometer onboard LPF. The two beams interfering in this interferometer are shown in red (measurement beam) and blue (reference beam). The measurement beam goes out of the optical bench setup passing a window (WIN) twice and is reflected by the freely falling test masses inside the electrode housings not shown here. This figure presents the principle setup of the LPF optical bench. Please note that the component sizes might slightly deviate from scale in this image. }
  \label{fig:LPFifo_x12}
\end{figure}

\gls{LPF} \cite{McNamara2008,Armano2018,Armano2016} was the technology demonstrator mission for the \gls{LISA} \cite{Danzmann2011,elisa13ARXIV,LISAMission}.
Launched in December 2015, it was in operation mode from March 2016 until its shutdown in July 2017.
The main scientific measurement of \gls{LPF} concerned the relative acceleration of its two hosted cubic test masses in free fall. 
These were situated in electrode housings at two opposite sites of an optical bench. 
The interferometer onboard \gls{LPF} measuring the distance changes and likewise the relative accelerations of the test masses, was the x12-interferometer, see Fig.~\ref{fig:LPFifo_x12}.
Since one of the interfering beams (red in this figure) was reflected at each of the test masses, any angular or lateral jitter of either the \gls{S/C} or the test masses coupled into the scientific measurement. 
We refer to this noise as \gls{TTL} coupling \cite{G21,NG21}. Without suppression, it was the most significant noise source between 20 and 200\,mHz \cite{Armano2018,Armano2016,Wanner2017}.

Two strategies were applied to remove the \gls{TTL} coupling from the readout and to consolidate the original noise model:
realignments of the test masses for TTL noise suppression and the subtraction of the noise by a dedicated fit model \cite{Wanner2017,dlr2020}.

The realignment method alone was not sufficient since the applied realignments relying on the analytical TTL models available during the mission did reduce but not fully mitigate the coupling noise. 
On the other hand, the subtraction method relied on an a-priori low magnitude of \gls{TTL} coupling since the subtraction added sensing noise at higher frequencies otherwise \cite{Wanner2017,HartigPhD}.

Combined, both TTL mitigation strategies successfully removed the \gls{TTL} noise. 
However, the underlying mechanisms leading to the observed cross-coupling were insufficiently modelled.
Also, the model fitted for \gls{TTL} subtraction did not resolve the physical dependency of the jitter coupling on the alignment parameters. 
Additionally, the fit had to be performed repeatedly to account for environmental changes and long-term drifts. 
On this regard, we continued investigating the \gls{TTL} coupling problem after the end of the \gls{LPF} mission and evaluated an analytical model that now describes the \gls{TTL} noise in \gls{LPF} \cite{LPFdata22}.
It will be presented in the following sections.

Beyond \gls{LPF}, \gls{TTL} coupling is a major noise source in space-based interferometers \cite{Troebs2018,Sasso2018_far-field,Sasso2018_misalignment,Wegener2020}.
Particularly in the case of \gls{LISA}, this noise source is widely discussed. To minimise the \gls{TTL} noise in the \gls{LISA} measurement band, it is planned to be suppressed by design, alignment and subtraction in post-processing. 
In that respect, a successful modelling of the \gls{TTL} noise in \gls{LPF} is of particular interest. By this, we do not only validate the planned suppression strategies but also boost our confidence in their effective application.

Within this paper, %
we evaluate an updated \gls{TTL} coupling model for \gls{LPF}.
We start with explaining our modelling strategy and define it from the general \gls{TTL} coupling derivations in \cite{G21,NG21} in Sec.~\ref{sec:TTLmodel_intro}.
The resulting analytical model, is presented in Sec.~\ref{sec:TTLmodel_LPSmodel}.
There, we introduce the different contributing \gls{TTL} coupling mechanisms, which are the lever arm coupling, the piston coupling and the coupling effects due to transmissive components.
We also distinguish between geometric an non-geometric \gls{TTL} coupling.
The analytical model is being verified against numerical simulations in Sec.~\ref{sec:TTLmodel_Verification}.
Furthermore, we characterise the stability of the coupling coefficients to uncertainties of the input parameters in (Sec.~\ref{sec:TTLmodel_Errors}).
In Sec.~\ref{sec:TTLmodel_Deltag}, we transform the presented analytical description of the interferometric length signal changes into a model describing the \gls{TTL} effects on the $\Delta g$ measurements in \gls{LPF}.
This transformation sets the base for a \gls{TTL} coupling analysis on the \gls{LPF} data \cite{LPFdata22}.
In Sec.~\ref{sec:discussion}, we discuss our findings and highlight the differences between our updated model and the \gls{TTL} noise equations used at the time of the mission.
All results will be summarised in the final section (Sec.~\ref{sec:summary}).

\section{General Assumptions and Modelling Strategy}
\label{sec:TTLmodel_intro}

For the analytical derivation of the \gls{TTL} coupling in \gls{LPF}, we have interpreted the coupling effects described in \cite{G21,NG21} for the \gls{LPF} setup. 
The distance changes of the two test masses were there captured by the x12-interferometer, see Fig.~\ref{fig:LPFifo_x12}.
It processes the differential phase changes of the interfering stable reference beam (blue) and the measurement beam (red) after its reflection at both test masses.

Several aspects of the \gls{TTL} mechanisms on the \gls{LPF} optical setup and the computation scheme presented here have already been addressed in \cite{G21,NG21}.
In \gls{LPF}, the \gls{TTL} coupling originated primarily from the lateral and angular jitter of the \gls{S/C} itself. The individual jitter of the two hosted test masses played a minor role.
For the analytic description of the effect, the \gls{S/C} jitter can be interpreted as a simultaneous jitter of the test masses relative the \gls{S/C}'s centre of mass and therefore refers to the setup with jittering test masses introduced in \cite{G21,NG21}.
The occurring \gls{TTL} coupling mechanisms split into the lever arm, the piston and the transmissive components' effect.

Furthermore, the two interfering beams in the x12-interferometer are assumed to be perfect Gaussian beams. 
This condition holds since both beams are local and little clipping and only weak ghost beams can be expected. Therefore, the residual imperfection of the beams' Gaussian shapes is assumed to be negligible.
Thus, the computation of the non-geometric part of the \gls{TTL} coupling in LISA Pathfinder follows the method explained in \cite{Wanner2014,NG21}.

For our evaluation, we chose to analyse the signal changes measured by the A-diode of this interferometer (PD12A).
The signals of the A- and B-diode differ due to the number of transmissions of the measurement beam, the slightly longer lever arm in the case of the B-diode and small differences of the beam alignments at the respective detector surface.
However, when considering only \gls{S/C} jitter, these differences are small since the alignment angle of the beam after its reflection at the second test mass stays unchanged. Thus, the only difference in the final signal originates from the small tilt-dependent lateral shift of the beam path multiplying with the nominal beam alignment angles at the detectors.
Due to the small differences of the coupling in both detectors, we restrict our analysis to the A-diode. 

Despite the agreement of several preconditions of the \gls{LPF} setup and the analysis in \cite{G21,NG21}, the analytical \gls{TTL} coupling model for \gls{LPF} cannot simply be evolved by inserting the system parameters into the presented equations.
The \gls{LPF} setup is more complicated in the sense that it hosts two test masses. 
As the measurement beam reflects first at the first test mass (TM1), any jitter of this test mass will generate a beam walk on the second test mass (TM2) and also changes the angle of incidence there. 
Consequently, the \gls{TTL} effects due to the jitters of both test masses are not independent of each other.
The analysis obtains additional complexity from both test masses being nominally tilted (about their centre of mass).
Therefore, the expected geometric path length changes of the measurement beam and its beam walk on the detector was computed independently from \cite{G21,NG21} in Mathematica \cite{mathematica}.
The computation assumes the lever arm and piston effect of both test masses and the transmission of the rotated measurement beam through the windows in between the test masses and the optical bench.

\begin{figure}
  \centering
  \includegraphics[width=0.5\columnwidth]{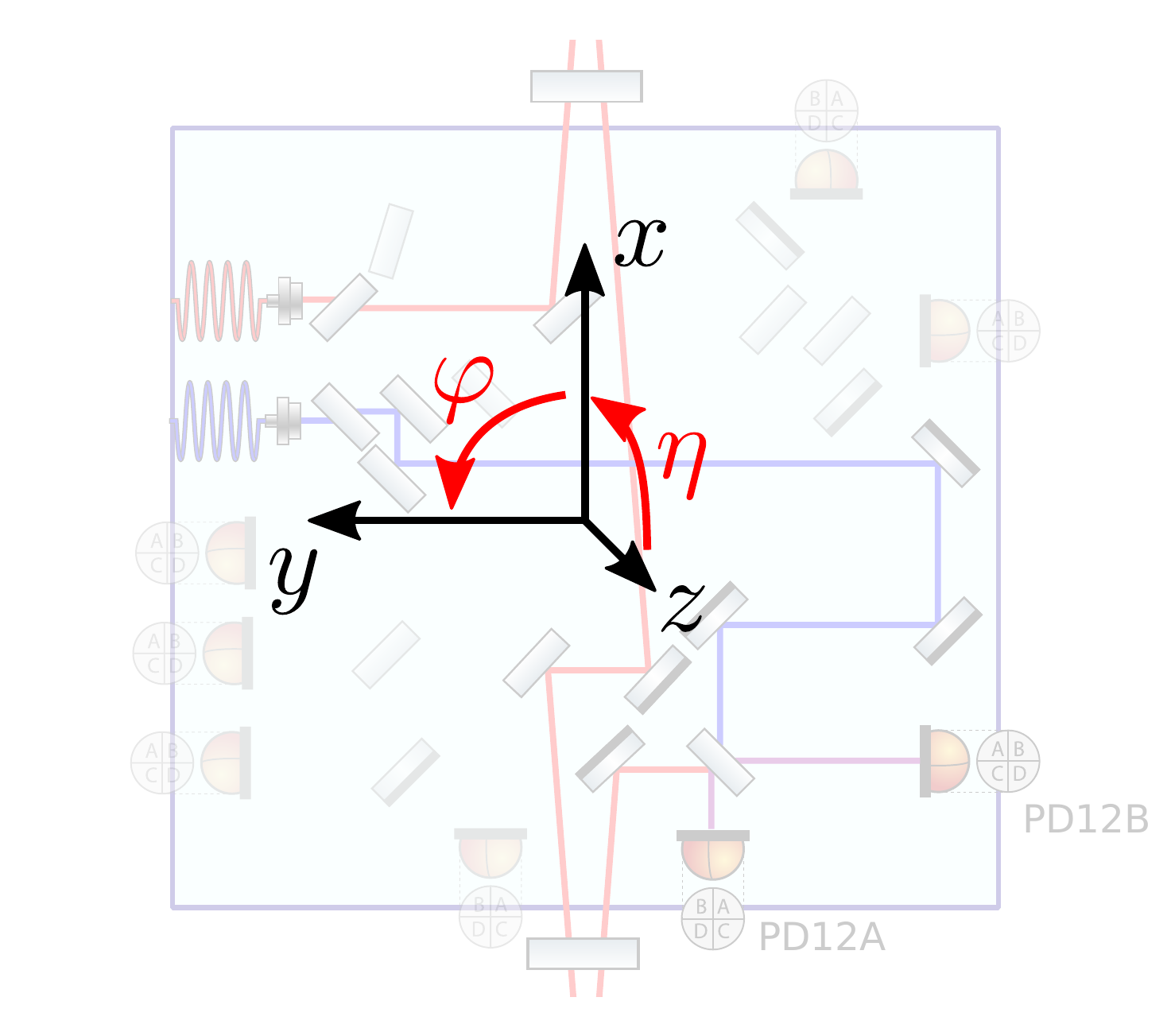}
  \caption{The global coordinate system that we use within this paper.
  Its centre is defined as the geometrical centre of the (non-shifted) optical bench. 
  The $xy$-plane is plane-parallel to the (non-rotated) optical bench surface and the $z$-axis points upwards.
  Accordingly, all yaw angles ($\varphi$) describe in-plane rotations, while the pitch angles ($\eta$) describe out-of-plane rotations.
  The coordinate system is defined with respect to the nominal location and orientation of the optical bench, but not fixed to it: any S/C jitter would change the optical bench alignment but not the coordinate system.}
  \label{fig:LPF_cs}
\end{figure}

The resulting equations are very complex due to the number of parameters included. 
Thus, we will present in the following sections only the series expression with inserted parameters (see App.~\ref{sec:parameters}).
Thereby, we use the coordinate system shown in Fig.~\ref{fig:LPF_cs}.
The full equations are shown in \cite{mathematica}.

\section{Modelling of the Longitudinal Path Length Signal Changes in LISA Pathfinder}
\label{sec:TTLmodel_LPSmodel}

We present in this section the \gls{TTL} coupling induced changes of the \gls{LPS} signal \cite{NG21,Wanner2014} at the A-diode of the x12-interferometer (Fig.~\ref{fig:LPFifo_x12}).
This \gls{S/C} jitter noise splits up into the lever arm and the piston effect \cite{G21,NG21}. 
Additionally, the windows between the optical bench and the vacuum housing of the test masses add coupling \cite{G21,NG21}, which is, in the case of \gls{LPF}, smaller than the lever arm and the piston effect.

\subsection{The Lever Arm Effect}
\label{sec:TTLmodel_LPSmodel_lever}

The lever arm effect describes the signal changes due to the tilt of the beam axis after the reflection at either of the test masses \cite{G21}. This tilt scales with the path length of the beam in between the two test masses and between the second test mass and the detector.
Furthermore, this tilt induces a beam walk on the photodiode surface which adds non-geometric TTL coupling \cite{NG21}.
We expand the resulting analytical equation in the tilt angles, insert the parameters provided in App.~\ref{sec:parameters} and discard negligible terms.
This yields
\begin{align}
\begin{split}
\text{LPS}_\text{lever}^\text{LPF} &\approx 
  45\cdot10^{-6}\frac{\text{m}}{\text{rad}}\,\varphi_\text{SC}
 -10\cdot10^{-6}\frac{\text{m}}{\text{rad}}\,\eta_\text{SC} \\
&+0.747\frac{\text{m}}{\text{rad}^2}\,\hat\varphi_1\,\varphi_\text{SC}
 +0.711\frac{\text{m}}{\text{rad}^2}\,\hat\varphi_2\,\varphi_\text{SC} \\
&+0.746\frac{\text{m}}{\text{rad}^2}\,\hat\eta_1\,\eta_\text{SC}
 +0.702\frac{\text{m}}{\text{rad}^2}\,\hat\eta_2\,\eta_\text{SC} \\
&-0.729\frac{\text{m}}{\text{rad}^2}\,\varphi_\text{SC}^2
 -0.724\frac{\text{m}}{\text{rad}^2}\,\eta_\text{SC}^2 \,.
\end{split}
\label{eq:TTL_lever}
\end{align}
The lever arm effect in \gls{LPF} induces linear and quadratic \gls{TTL} coupling in the \gls{S/C} angles $\varphi_\text{SC}$ (in-plane / yaw jitter) and $\eta_\text{SC}$ (out-of-plane / pitch jitter).
The coupling depends on the absolute test mass alignment angles: $\hat\varphi_i$,\,$\hat\eta_i$, where $i\in\{1,2\}$ indicates the considered test mass.
See Fig.~\ref{fig:LPF_cs} for the definition of the coordinate system and the orientation of the angles.
It is $\hat\varphi_i=0$, if the reflecting test mass surface is parallel to a plane including the $y$-axis. The same applies to $\hat\eta_i=0$ and the $z$-axis.
All cross-plane effects have shown to be negligible in the lever arm case. 

\subsection{The Piston Effect}
\label{sec:TTLmodel_LPSmodel_piston}

The piston effect covers the \gls{LPS} signal changes due to the translation of the measurement beam's reflection points at the test masses by the motion of their reflective surfaces. 
For angular \gls{S/C} jitter, the surface motion into the beam path originates from the offset of the centre of rotation from the beam's reflection point \cite{G21,NG21}.
Furthermore, the lateral jitter of the \gls{S/C} shifts the surface into the beam paths if they are nominally tilted with respect to the optical bench.
It splits into the in-plane jitter $y_\text{SC}$ (i.e.\ parallel to the $y$-axis in Fig.~\ref{fig:LPF_cs}) and the (vertical) out-of-plane jitter $z_\text{SC}$ (i.e.\ parallel to the $z$-axis in Fig.~\ref{fig:LPF_cs}).
We found
\begin{align}
\begin{split}
\text{LPS}_\text{piston}^\text{LPF} &\approx 
  19\cdot10^{-6}\frac{\text{m}}{\text{rad}}\,\varphi_\text{SC}
 +14\cdot10^{-6}\frac{\text{m}}{\text{rad}}\,\eta_\text{SC} \\
&+1.994\frac{1}{\text{rad}}\,\left[(-\hat\varphi_1+\hat\varphi_2)\,y_\text{SC}+(\hat\eta_1-\hat\eta_2)\,z_\text{SC} \right] \\
&-0.318\frac{\text{m}}{\text{rad}^2}\,\hat\varphi_1\,\varphi_\text{SC}
 -0.340\frac{\text{m}}{\text{rad}^2}\,\hat\varphi_2\,\varphi_\text{SC} \\
&-0.319\frac{\text{m}}{\text{rad}^2}\,\hat\eta_1\,\eta_\text{SC}
 -0.339\frac{\text{m}}{\text{rad}^2}\,\hat\eta_2\,\eta_\text{SC} \\
&+0.010\frac{\text{m}}{\text{rad}^2}\,(-\hat\varphi_1+\hat\varphi_2)\,\eta_\text{SC} \\
&+0.329\frac{\text{m}}{\text{rad}^2}\,\varphi_\text{SC}^2
 +0.329\frac{\text{m}}{\text{rad}^2}\,\eta_\text{SC}^2 \,.
\end{split}
\label{eq:TTL_piston}
\end{align}
Here, we see not only linear and quadratic \gls{TTL} coupling terms but also a small cross-plane term depending on the yaw alignment of the test mass and the pitch jitter of the \gls{S/C}.
It depends on the vertical offset of the \gls{S/C}'s centre of mass from the beam's reflection points and their in-plane incoming angles at the test masses.
There is no analogue cross-plane term for yaw jitter since the multiplicative lateral offset is one order of magnitude and the out-of-plane incoming angles are three orders of magnitude smaller (Tabs.~\ref{tab:parameters_setup}, \ref{tab:parameters_angles}).
For details, compare with \cite[Eq.~(33)]{G21} or \cite{mathematica}.
Like the lever arm effect, the piston effect depends on the nominal angular alignment of the test masses. Their lateral alignment, however, does not couple with the \gls{S/C} jitter \cite{HartigPhD}.

\subsection{Refraction at the Windows}
\label{sec:TTLmodel_LPSmodel_window}

Along their path in between the test masses and between the second test mass and the detector, the (tilted) beam passed the windows of the vacuum chambers of the electrode housings (see Fig.~\ref{fig:LPFifo_x12}).
The laser beam refracts when entering and exiting the window. 
The propagation distance and direction inside the material depends on the beam's angle of incidence on the window. 
Since this incidence angle varies if the \gls{S/C} jitters, the latter yielded additional \gls{TTL} coupling by the window.
It is described by
\begin{align}
\begin{split}
\text{LPS}_\text{window}^\text{LPF} &\approx 
 -0.018\frac{\text{m}}{\text{rad}^2}\,\left[\hat\varphi_2\,\varphi_\text{SC} +\hat\eta_2\,\eta_\text{SC}\right] \\
&+0.009\frac{\text{m}}{\text{rad}^2}\,\varphi_\text{SC}^2
 +0.009\frac{\text{m}}{\text{rad}^2}\,\eta_\text{SC}^2 \,.
\end{split}
\label{eq:TTL_window}
\end{align}
By comparing Eq.~\eqref{eq:TTL_window} with Eqs.~\eqref{eq:TTL_lever} and~\eqref{eq:TTL_piston}, we see that the \gls{TTL} contribution of the windows is smaller than the lever arm and the piston effect. 
We compare the three presented \gls{TTL} coupling contributors in Sec.~\ref{sec:TTLmodel_Mechanisms}.

\subsection{The Full TTL Coupling in LPF}
\label{sec:TTLmodel_LPSmodel_full}

As stated above, the \gls{TTL} coupling in \gls{LPF} can be described via the lever arm effect, the piston effect and the signal changes due to the transmissive components along the beam path. 
The full \gls{TTL} coupling noise is approximately the sum of the respective Eqs.~\eqref{eq:TTL_lever}-\eqref{eq:TTL_window}. Only small deviation appear when computing the combined effect:
The full geometric \gls{TTL} coupling can simply be evaluated by adding the geometric contributions of the three effects.
The non-geometric \gls{TTL} coupling is then computed as described in \cite{NG21} by taking into account the beam walk on the photodiode accumulated due to the three coupling mechanisms (sum of the beam walks due to the three effects).
Since the beam walk appears squared in the non-geometric formalism, cross-terms of the single coupling effect add in the full analytical \gls{TTL} coupling model. 
Despite these cross-terms and hence the deviation of the final result from the sum of the equations presented above is small, we state here the model computed for full, combined effect for correctness:
\begin{align}
\begin{split}
\text{LPS}^\text{LPF} &\approx 
  63\cdot10^{-6}\frac{\text{m}}{\text{rad}}\,\varphi_\text{SC}
 +5\cdot10^{-6}\frac{\text{m}}{\text{rad}}\,\eta_\text{SC} \\
&+1.994\frac{1}{\text{rad}}\,\left[(-\hat\varphi_1+\hat\varphi_2)\,y_\text{SC}+(\hat\eta_1-\hat\eta_2)\,z_\text{SC} \right] \\
&+0.419\frac{\text{m}}{\text{rad}^2}\,\hat\varphi_1\,\varphi_\text{SC}
 +0.362\frac{\text{m}}{\text{rad}^2}\,\hat\varphi_2\,\varphi_\text{SC} \\
&+0.417\frac{\text{m}}{\text{rad}^2}\,\hat\eta_1\,\eta_\text{SC}
 +0.354\frac{\text{m}}{\text{rad}^2}\,\hat\eta_2\,\eta_\text{SC} \\
&+0.010\frac{\text{m}}{\text{rad}^2}\,(-\hat\varphi_1+\hat\varphi_2)\,\eta_\text{SC} \\
&-0.390\frac{\text{m}}{\text{rad}^2}\,\varphi_\text{SC}^2
 -0.385\frac{\text{m}}{\text{rad}^2}\,\eta_\text{SC}^2 \,.
\end{split} 
\label{eq:TTL_full}\\
\begin{split}
&\equiv C_\varphi\,\varphi_\text{SC} + C_\eta\,\eta_\text{SC} + C_y\,y_\text{SC} + C_z\,z_\text{SC} \\
&+ \mathcal{O}(\varphi^2,\eta^2) \,,
\end{split}
\end{align}
where $C_i$, $i\in\{\varphi,\eta,y,z\}$, are the coupling coefficients scaling the linear \gls{S/C} jitter in the respective degree of freedom.

Note that solely non-geometric effects that depend on the three coupling mechanisms presented above add to the signal for \gls{S/C} jitter.
Further, independent non-geometric coupling contributions presented in \cite{NG21}, e.g.\ nominal offsets between the interfering beams, are not present here. These would only couple with a dynamic change in the measurement beam's incidence angle at the detector. 
However, any \gls{S/C} jitter-induced changes in the direction of the measurement beam after its reflection at the first test mass will cancel after its reflection at the second test mass. 
This is because, from the perspective of the \gls{S/C}, both test masses jitter synchronously about its centre of rotation.
Thus, the angular alignment of the measurement beam at the detector remains constant.

In general, we find by Eq.~\eqref{eq:TTL_full}
that the lateral and angular \gls{TTL} coupling in \gls{LPF} depended strongly on the alignment of the test masses.
Second-order coupling occurs only due to angular \gls{S/C} jitter in the respective two degrees of freedom.

We show in Sec.~\ref{sec:TTLmodel_Verification} that Eq.~\eqref{eq:TTL_full} is consistent with numerical simulations.
Furthermore, we can, for the first time reconstruct the performance changes due to test mass realignments during the \gls{LPF} mission \cite{LPFdata22}.
In addition, the coupling coefficients computed with Eq.~\eqref{eq:TTL_full} for mission data of \gls{TTL} of a particularly designed coupling experiment match in almost all cases their fitted counterparts within their error bars (see \cite{LPFdata22} for further details.

\subsection{TTL Noise Minimisation}
\label{sec:TTLmodel_Minimisation}
Eq.~\eqref{eq:TTL_full} shows that the degree of \gls{S/C} jitter coupling into the \gls{LPS} signal in \gls{LPF} depended strongly on the angular alignment of the test masses.
By minimising Eq.~\eqref{eq:TTL_full} regarding the test mass angles $\hat\varphi_{1,2}$ and $\hat\eta_{1,2}$, alignments can be defined that fully cancel the linear \ac{TTL} coupling due to \ac{S/C} jitter.

However, from Eq.~\eqref{eq:TTL_full} we cannot gain an fixed alignment angles valid for the full mission duration:
First, as will show in Sec.~\ref{sec:TTLmodel_Errors}, the test mass independent \gls{S/C} jitter terms (first row of Eq.~\eqref{eq:TTL_full}) depend highly on the stability of the optical setup and could therefore change during the mission.
Second, the measurement of the test masses' angular alignment changes was highly precise in \gls{LPF}, while their true absolute alignments (tilt with respect to the global coordinate system, see Fig.~\ref{fig:LPF_cs}) were uncertain.
Thus, test mass realignment angles would have to be found that counteract both the unknown initial alignment and the test mass independent linear coupling terms.

Within this paper, we define as the nominal case, the angular test mass alignment that minimises the differential angle of the measurement and the reference beam at the detectors.
Precisely, the \gls{DWS} \cite{Wanner2012,Morrison1994} angles of the interfering beams in the x1-interferometer (measuring the relative alignment between the \gls{S/C} and the first test mass) and the x12-interferometer (Fig.~\ref{fig:LPFifo_x12}).
These angles are summarised in Tab.~\ref{tab:IfoCADangles}.
They have been computed based on the measurements of the optical setup of the \gls{LPF} flight model \cite{TNoptocad}.

\begin{table}
  \caption{Nominal angular alignments of the test masses. 
   These angles have been defined to minimise the the DWS angles measured by the x1- and x12-interferometer.}
\label{tab:IfoCADangles}
\begin{tabular}{ccc}
  \toprule
  angle & [unit] & nominal realignment \\
  \midrule
  $\hat\varphi_1$ & [$\upmu$rad] &   7.98 \\
  $\hat\varphi_2$ & [$\upmu$rad] & -60.7 \\
  $\hat\eta_1$    & [$\upmu$rad] &  13.0 \\
  $\hat\eta_2$    & [$\upmu$rad] &   0.355 \\
  \bottomrule
\end{tabular}
\end{table}

\subsection{Comparison of the Noise Contributors}
\label{sec:TTLmodel_Mechanisms}

\begin{figure}
  \includegraphics[width=\columnwidth]{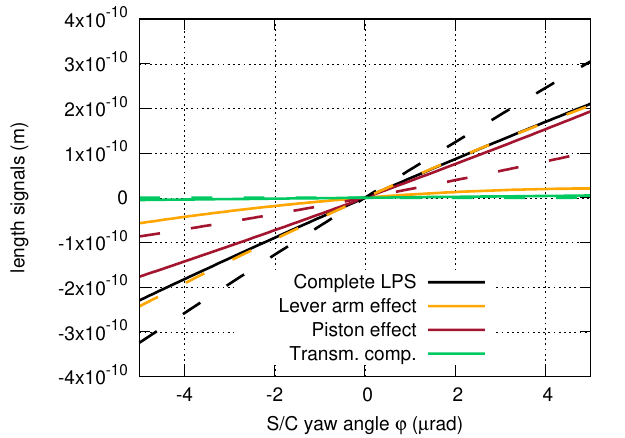} \\
  \includegraphics[width=\columnwidth]{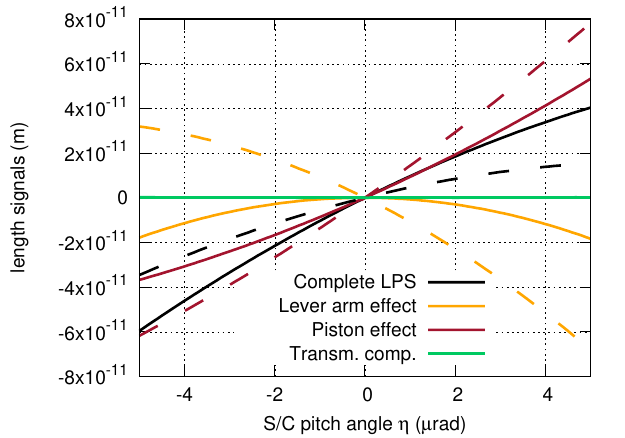}
  \caption{The TTL coupling for S/C rotations in yaw (top) and pitch (bottom) and its contributors: the lever arm effect, the piston effect and the transmissive components. 
  Dashed curves: coupling if all test mass alignment angles are set to zero.
  Solid curves: coupling considering nominal test mass alignments (Tab.~\ref{tab:IfoCADangles}).
  All curves have been derived analytically.
  We see that the coupling strongly depends on the test mass alignment.
  The lever arm and the piston effect are the most significant noise contributors.}
  \label{fig:LPSandContributors}%
\end{figure}

In \gls{LPF}, the \gls{TTL} coupling originated from both angular and lateral \gls{S/C} jitter \cite{LPFdata22}. The lateral jitter contributed significant linear coupling noise. Its magnitude depended only on the angular test mass alignment and the beam orientation before the reflection \cite{G21,mathematica}. 
In the case of angular jitter coupling, all three \gls{TTL} coupling mechanisms presented above add to the full measured \gls{TTL} noise. 
Their contribution to the final signal is visualised in Fig.~\ref{fig:LPSandContributors}. %
We find that the lever arm and the piston effect are the dominating mechanisms for angular \gls{S/C} jitter.
The lever arm effect cannot be fully mitigated by design.
Consequently, a direct \gls{TTL} coupling suppression (i.e.\ by realignment) would only have been possible, when the lever arm and the (angular) piston effect would counteract each other. 

\begin{figure}
  \includegraphics[width=\columnwidth]{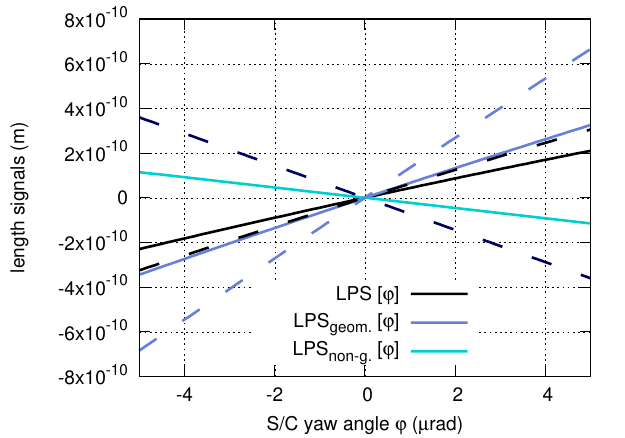} \\
  \includegraphics[width=\columnwidth]{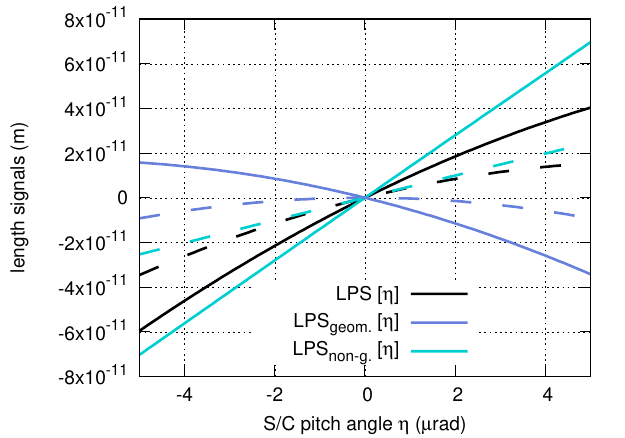}
  \caption{Comparison of the contributions of the geometric and the non-geometric coupling to the full LPS signal in the case of yaw ($\varphi$) and pitch ($\eta$) rotations.
  Dashed curves: coupling if all test mass alignment angles are set to zero.
  Solid curves: coupling considering nominal test mass alignments (Tab.~\ref{tab:IfoCADangles}).
  All signals have been derived analytically.
  For both rotations, we see that the geometric as well as the non-geometric coupling terms are relevant for describing the full TTL coupling in LPF.}
  \label{fig:LPSandContributors_LPSvsOPD}
\end{figure}

Besides the contributions of the single \gls{TTL} coupling mechanisms, we compare here the geometric and non-geometric signal contributions. 
As shown in Fig.~\ref{fig:LPSandContributors_LPSvsOPD}, both add significant coupling to the full signal.
Therefore, neither can be neglected when analytically describing the observed \gls{TTL} coupling.

In general, the \gls{TTL} contributions depended highly on the test mass alignments. 
While for small \gls{S/C} jitter amplitudes (few nano-radians), the coupling can be linearised, the slope significantly changes if the test masses were rotated.
Therefore, it was essential to compute the series expansion of the \gls{LPS} signal up to second orders of the small angles.

\section{Verification of the Analytical Model}
\label{sec:TTLmodel_Verification}

To verify the presented analytical description of \gls{TTL} coupling (Eq.~\eqref{eq:TTL_full}), we compared it with numerical simulations.
This has been done via the the C++ library IfoCAD \cite{IfoCAD,Wanner2012,Kochkina2013}. 
All relevant components of the \gls{LPF} setup have been implemented in IfoCAD, i.e., all optical components of the LPF optical bench and the windows of the vacuum chambers surrounding the test masses. 
The implementation copies the in-flight model of \gls{LPF} \cite{TNoptocad}.
The two beams are then automatically traced through this setup and interfered at the detectors.

The angular alignment of the test masses in IfoCAD has been adjusted to the angles defined in Tab.~\ref{tab:IfoCADangles}, i.e., the angles minimising the \gls{DWS} angles in the x1- and x12-interferometer.

\begin{figure}
  \includegraphics[width=\columnwidth]{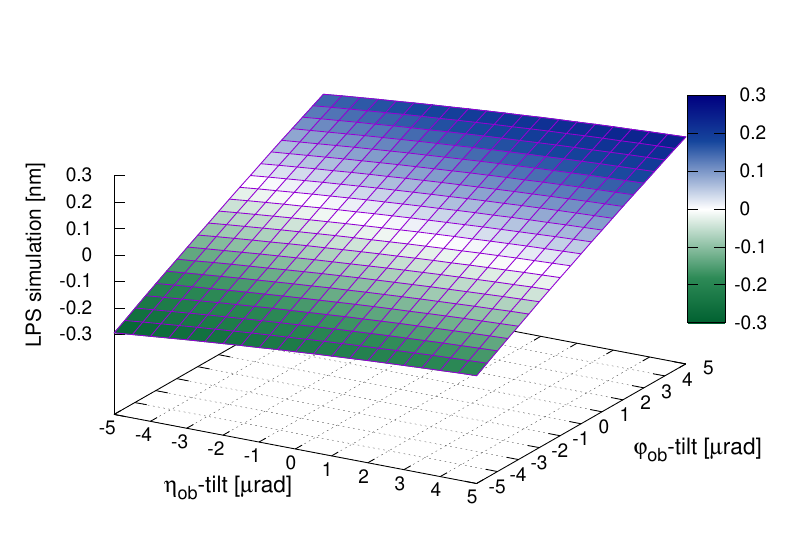} 
  \caption{The LPS computed by IfoCAD for S/C rotations around its centre of mass. The computation considers the nominal test mass alignments (Tab.~\ref{tab:IfoCADangles}).}
  \label{fig:LPSifo}
\end{figure}

The full \gls{TTL} coupling for angular variations is shown in Fig.~\ref{fig:LPSifo}. 
We plot the IfoCAD-derived coupling for angular variations up to 5\,$\upmu$rad.
This was the injected jitter amplitude of several experiments during the \gls{LPF} mission \cite{LPFdata22}.
The jitter in noise runs was significantly smaller.
The plot demonstrates that the coupling can be approximated by linear formalisms for small jitter amplitudes.

\begin{figure}
  \includegraphics[width=\columnwidth]{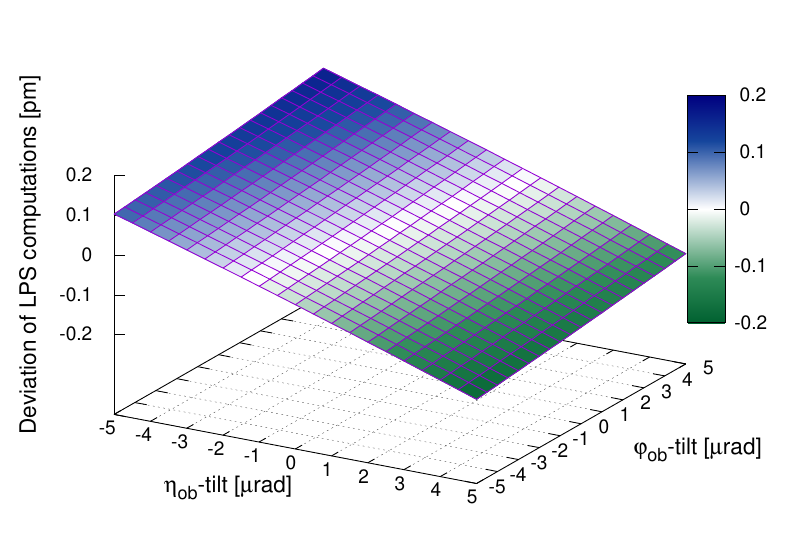} 
  \caption{The deviation between the analytically derived LPS and the LPS computed by IfoCAD for S/C rotations around its centre of mass. Both computations considers the nominal test mass alignments (Tab.~\ref{tab:IfoCADangles}). The residual is three orders of magnitude smaller than the signal itself. It originates from simplifying assumptions in the evaluation of the analytical non-geometric signal.}
  \label{fig:LPSaccuracy}
\end{figure}

In Fig.~\ref{fig:LPSaccuracy}, we compare our analytical model (Eq.~\eqref{eq:TTL_full}) against the IfoCAD result. 
The difference between both models is more than three orders of magnitude smaller than the original coupling. 
The residual difference originates from the series expansion of the analytical model and likely from simplifying assumptions in the computation of the non-geometric coupling contributions \cite{NG21}.
So, the deviations are more than one order of magnitude smaller for the sole geometric coupling or the comparison of the evaluated and simulated beam walk on the detector \cite{mathematica}.
The shown small deviation validates our analytical \gls{TTL} coupling description.


Lateral jitter coupling almost only originates from geometric \gls{TTL} coupling effects \cite{NG21}. A comparison between the analytical model and the simulated lateral jitter only showed deviation of the order of the machine accuracy \cite{HartigPhD}.

\section{Uncertainty of the Modelled Coefficients}
\label{sec:TTLmodel_Errors}

The analytical model presented in this manuscript relies on our best knowledge of the setup and beam parameters.
However, the measurement accuracy of the latter is technologically limited and the parameters can change during the mission due to stresses and relaxation of the materials. 
In general, we assume the following parameters to be subject to measurement inaccuracies or in-flight changes of the optical system.

\paragraph{The lever arm lengths:}
The lever arm lengths (compare setup parameters list in appendix, Tab.~\ref{tab:parameters_setup}) have been computed via IfoCAD. Their numbers rely on the accuracy of the position parameters of the optical components inserted in the script.
In flight, the lever arm lengths are affected by distortions of the optical bench itself but also by longitudinal displacements of the test masses. 
The corresponding length variations would change the angular coupling coefficients $C_\varphi$ and $C_\eta$. 
However, we assume the overall changes of the lever arm length to be several orders of magnitude smaller than its full length.
Since a 1\% length change of a lever arm approximately translates into 2\% change of the angular coupling coefficients, this coupling coefficient change would also be several magnitudes smaller than its absolute value.
Therefore, lever arm length uncertainties will be neglected in the following.

\paragraph{The longitudinal offsets between the reflection point at the test mass and the centre of rotation:}
\label{sec:TTLmodel_errors_dlong}
The measurement of the longitudinal offset between the beam's point of reflection at the test masses and the satellite's centre of mass is limited by the determination of the exact position of the latter.
The uncertainty of the \gls{LPF}'s centre of mass is $\pm$5\,mm in all axes \cite{TNdfacs}.
Furthermore, this centre varies over the mission time due to fuel (cold gas) consumption. 
Based on this consumption, we assume here a variation of the longitudinal parameter of the \gls{S/C}'s centre of mass of $-0.5\,\text{mm}$\,...\,$+2.3\,\text{mm}$ \cite{valerio}.
Additional changes of this offset, e.g. due to distortions of the optical bench, are assumed to be negligible.
The changes of the longitudinal offsets yield roughly a third of the errors summarised in the following (Tab.~\ref{tab:anaerrors}).

\begin{figure}
  \includegraphics[width=\columnwidth]{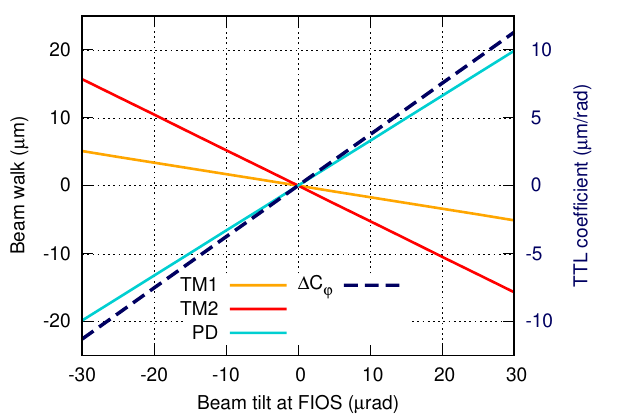} \\
  \includegraphics[width=\columnwidth]{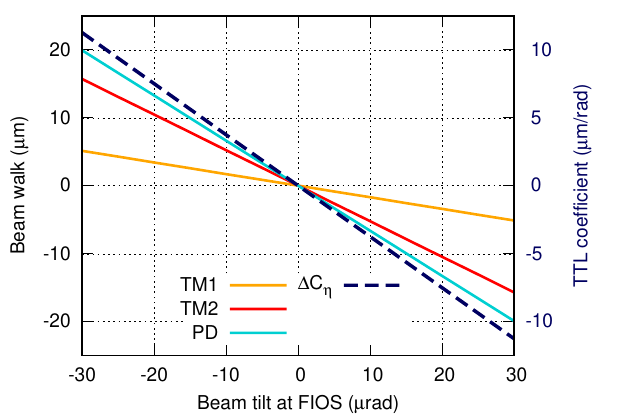}
  \caption{The beam walk on the two test masses and the photodiode (left axis) due to a yaw (top) and pitch (bottom) tilt of the beam at the FIOS, as well as the resulting TTL coupling coefficient variations (right axis). In the given range, the TTL coefficient $C_\eta$ changes by more than 10\,$\mu$m/rad from its nominal value. All numbers have been computed with IfoCAD.}
  \label{fig:TTL_fios} %
\end{figure}

\paragraph{The lateral offsets between the reflection point at the test mass and the centre of rotation:}
\label{sec:TTLmodel_errors_dlat}
Neither lateral shifts of the test masses nor of the \gls{S/C} have a measurable effect on the offset between the point of reflection and the centre of rotation. 
Thus, these offsets would only change if the path of the beam hitting the test masses differs from its simulation with IfoCAD.
Such distortions of the beam direction can result from thermal stresses on the optical system and significantly change the linear angular coupling terms (see Fig.~\ref{fig:TTL_fios}, further discussion in Para.~e).
In addition, the measurement of the lateral position of the \gls{S/C}'s centre of mass has an uncertainty of $\pm$5\,mm \cite{TNdfacs} and it further varies by $-0.5\,\text{mm}$\,...\,$+3.8\,\text{mm}$ along the $y$-axis and marginally along the $z$-axis due to the cold gas consumption \cite{valerio}. 
All these changes mostly affect the constant offset of the angular \gls{TTL} coupling coefficients. They couple with the test mass alignment only to a negligible degree. 

\paragraph{Window properties:}
We have shown in Sec.~\ref{sec:TTLmodel_Mechanisms} that the windows in between the optical bench and the test masses are small \gls{TTL} contributors compared to the lever arm and the piston effect. Therefore, small changes of their thickness, alignment or refractive index would be negligible in the full TTL coupling estimate.

\paragraph{The angular beam alignment:}
\label{sec:TTLmodel_errors_beamtilt}
Changes in the beam alignment at their source or due to distortions of the optical bench \cite{Killow2016,LPFlongterm22} would, on the one hand, change the lateral position of the reflection points on the test masses (see above). 
On the other hand, the angles of incidence at the test masses change, which yields a small change of the piston effect.
In general, these changes couple linearly with the \gls{S/C} angular jitter.
However, in-flight changes of the beam alignment due to stresses and relaxations of the optics would be (partially) corrected by test mass rotation due to the \gls{DFACS} \cite{Schleicher2018,Armano2019_stability} control loop, which keeps the \gls{DWS} readout \cite{Wanner2012,Morrison1994} of the relative beam alignments at the photodiodes stable.
Due to the number of potential origins of beam misalignments, their actual effect cannot be easily determined mathematically and we neglect it in the overall analysis.
For the special case of a beam rotation at the \gls{FIOS} (e.g.\ thermally induced \cite{Killow2016}), we exemplary show the effect on the coupling coefficients (uncorrected by test mass alignments) in Fig.~\ref{fig:TTL_fios}. %
We see that angular beam alignment changes can have a significant effect on the linear angular \gls{TTL} coupling coefficients.

\paragraph{The beam parameters:}
\label{sec:TTLmodel_errors_bparam}
The measurement accuracy of the waist size and the distance from waist of both interfering beams is limited. The resulting uncertainties yield a possible error of the non-geometric coupling estimate.
According to \cite{Killow2016}, the measurement of the beam's waist size was very accurate ($w_{0m} = (542 \pm 4)\,\upmu$m, $w_{0r} = (500 \pm 8)\,\upmu$m) yielding a maximal error of the Rayleigh range of 3.4\%. 
The measurement of the distance between the waist and the beam source ($z_m = (142 \pm 19)$\,mm, $z_r = (500 \pm 8)$\,mm) corresponds to a measurement error of 3.6\% for the distance from waist at the diode PD12A.

\paragraph{Arbitrary wavefront or detector errors:}
Arbitrary imperfections of the beams' wavefronts and the detector surface alter the measured phase signal. These effects are not included in the analytical model but can cause a small deviation between the analytically predicted coupling and the photodiode readout.
\\

To determine the minimal and maximal deviations of the coupling coefficients, we have to insert the parameter variations characterised above into the extended exact coupling equations.
We restrict this analysis to the variations that we clearly characterised with numbers (summarised in Tab.~\ref{tab:parameters_uncertainties}).
See \cite{mathematica} for further details about the computation.
The described parameter changes or measurement inaccuracies mainly affect the angular coupling coefficients (i.e.\ multipliers of $\varphi_\text{SC}$ and $\eta_\text{SC}$).
The lateral jitter coupling is mostly described by geometric \gls{TTL} coupling \cite{G21} and strongly depends on the angular test mass alignment. 
Any other small changes of the optical setup have a negligible effect on the lateral \gls{TTL} coefficients (i.e.\ multipliers of $y_\text{SC}$ and $z_\text{SC}$).

\begin{table}
\caption{Uncertainties of the analytical coefficients assuming errors in the measurement of the beam parameters and the centre of mass of the S/C.}
\begin{tabular}{l|ccc}
\toprule
term & coeff.\ [m/rad$^2$] 
  & abs.\,error [m/rad$^2$] & rel.\,error [\%] \\
\midrule
$\varphi_\text{SC}\,\hat\varphi_1$ & 0.419
  & -0.034/+0.031 & -8.2/+7.5 \\
$\varphi_\text{SC}\,\hat\varphi_2$ & 0.362
  & -0.037/+0.039 & -10.0/+10.9 \\
\midrule
$\eta_\text{SC}\,\hat\eta_1$ & 0.417
  & -0.033/+0.030 & -8.0/+7.2 \\
$\eta_\text{SC}\,\hat\eta_2$ & 0.354
  & -0.035/+0.038 & -9.9/+10.8 \\
\bottomrule
\end{tabular}
\label{tab:anaerrors}
\end{table}

\begin{table}
  \caption{The parameter uncertainties that we assume in our analytical model.
  The corresponding nominal values are given in Tabs.~\ref{tab:parameters_setup} and \ref{tab:parameters_beam}.
  The points of reflection (PoR) are defined with respect to the \gls{S/C} centre of mass.}
  \begin{tabular}{p{0.34\textwidth}@{\hskip 8pt}r}
  \toprule
  parameter & uncertainty \\
  \midrule 
  longitudinal distance between the PoR at TM1 and the S/C centre of mass & $-$7.3/+5.5\,mm \\
  lateral distance between the PoR at TM1 and the S/C centre of mass     & $-$5.5/+8.8\,mm \\
  vertical distance between the PoR at TM1 and the S/C centre of mass    & $\pm$5\,mm \\
  longitudinal distance between the PoR at TM2 and the S/C centre of mass &  $-$7.3/+5.5\,mm \\
  lateral distance between the PoR at TM2 and the S/C centre of mass     & $-$5.5/+8.8\,mm \\
  vertical distance between the PoR at TM2 and the S/C centre of mass    & $\pm$5\,mm \\
  \midrule
  Rayleigh range of the measurement beam & $\pm$0.014\,m \\
  Rayleigh range of the reference beam   & $\pm$0.025\,m \\
  distance from waist of the measurement beam & $\pm$0.019\,m \\
  distance from waist of the reference beam   & $\pm$0.008\,m \\
  \bottomrule
  \end{tabular}
\label{tab:parameters_uncertainties}
\end{table}

For the test mass alignment dependent terms of the angular \gls{TTL} coupling coefficients we show the computed errors in Tab.~\ref{tab:anaerrors}.
These are of particular interest for the analysis of the \gls{TTL} coupling during the \gls{LPF} mission since angular realignments of the test masses have been applied for \gls{TTL} coupling suppression.
Thus, the errors in Tab.~\ref{tab:anaerrors} show the accuracy of the estimate of the test mass alignment dependency of the coupling noise.
As discussed in the paragraphs above, our knowledge of the position of the \gls{S/C}'s centre of mass and the beam parameters are the largest contributors to these terms. 

We intentionally do not show here the uncertainties of the linear coefficients, that are independent of the test mass alignments (i.e.\ the first two terms in Eq.~\eqref{eq:TTL_full}).
In \gls{LPF}, we did not have access to the exact test mass alignment angles ($\hat\varphi_i=\varphi_{0i}+\varphi_{i},\hat\eta_i=\eta_{0i}+\eta_{i}$) but only to their changes ($\varphi_i,\eta_i$).
Thus, Eq.~\eqref{eq:TTL_full}) can be split in
\begin{align}
\begin{split}
 &\,\text{LPS}^\text{LPF}(\varphi_{0i}+\varphi_{i},\eta_{0i}+\eta_{i}) \\
=&\,\text{LPS}^\text{LPF}(\varphi_{i0},\eta_{i0})
   +\text{LPS}^\text{LPF}(\varphi_i,\eta_i)
\end{split}
\label{eq:TTL_full_0}
\end{align} 
The unknown part of the test mass angle dependent coupling (first term in Eq.~\ref{eq:TTL_full_0}) would be assigned to the (no longer) test mass independent linear coefficient. Their sum form the `constant offsets' of the linear coupling coefficients. 
The additive term is approximately of the same magnitude as the other linear terms. 
Therefore, the exact value of the constant offset of the linear coupling terms cannot be determined in practice.

\section{Tilt-To-Length Noise in the $\Delta g$ Measurements}
\label{sec:TTLmodel_Deltag}

The analytical modelling presented so far, showed the \gls{TTL} coupling contribution to the length signal measured at the diode PD12A.
To prove the principle of laser interferometric gravitational wave measurements in space with \gls{LPF}, this length signal has been transformed to an estimate of the distance changes of the test masses
\begin{align}
\Delta x = x_1-x_2 \,, 
\label{eq:x12_def}
\end{align}
which is achieved by the multiplication with the factor $1/(2\cos(\beta_{y2}))$. Correspondingly, the same correction is applied to the \gls{TTL} noise contribution.
Note that the sign convention in Eq.~\eqref{eq:x12_def} was chosen in accordance with the $o_{12}$ definition in \cite{Armaro2022_OMS}.

Furthermore, we take the second derivative of the \gls{TTL} coupling in the \gls{S/C} jitter variables. 
This yields the noise contribution to the $\Delta g$ measurement, which is the central parameter describing the stability of the test masses. 
Neglecting second- and higher-order terms,
we find
\begin{align}
\Delta g_\text{xacc}^\text{ana} = 
C_\varphi^\text{ana}\,\ddot{\varphi}_\text{SC}  %
+ C_\eta^\text{ana}\,\ddot{\eta}_\text{SC}  %
+ C_y^\text{ana}\,\ddot{y}_\text{SC}  %
+ C_z^\text{ana}\,\ddot{z}_\text{SC}  %
\label{eq:anamodel}
\end{align}
with 
\begin{align}
C_\varphi^\text{ana} =&\, C_{\varphi,0}
+0.210^{+0.017}_{-0.016}\,\frac{\mathrm{m}}{\mathrm{rad}^2} \,\varphi_{1}
+0.182^{+0.018}_{-0.020}\,\frac{\mathrm{m}}{\mathrm{rad}^2} \,\varphi_{2} \label{eq:anaCphi}\\
\begin{split}
C_\eta^\text{ana} =&\, C_{\eta,0}
+0.209^{+0.017}_{-0.015}\,\frac{\mathrm{m}}{\mathrm{rad}^2} \,\eta_{1}
 +0.177^{+0.018}_{-0.019}\,\frac{\mathrm{m}}{\mathrm{rad}^2}\,\eta_{2} \\
&+0.005^{+0}_{-0}\,\frac{\mathrm{m}}{\mathrm{rad}^2} \,(-\varphi_{1}+\varphi_{2})
\end{split} \label{eq:anaCeta}\\
C_y^\text{ana} =&\, C_{y,0}
+\, 1.000^{+0}_{-0} \ \frac{1}{\text{rad}}\,(-\varphi_{1}+\varphi_{2}) \label{eq:anaCy}\\
C_z^\text{ana} =&\, C_{z,0}
+\, 1.000^{+0}_{-0} \ \frac{1}{\text{rad}}\,(\eta_{1}-\eta_{2}) \,, \label{eq:anaCz}
\end{align}
where the offsets $C_{i,0},\,i\in\{\varphi,\eta,y,z\},$ depend on the setup parameters (App.~\ref{sec:parameters}). 
In the previous section, we touched upon the topic that these offsets significantly changed with parameter changes. 
However, the change rate was negligible in short time segments (days or a few weeks) without experiments.
Thus, we can assume the $C_{i,0}$ to be constant in the data analysis of single time segments.
Furthermore, since the exact test mass alignment remained unknown, the \gls{TTL} coupling depending on the nominal test mass angles would also be assigned to these coefficients in the analysis. 
The pitch and yaw angles in Eqs.~\eqref{eq:anaCphi}-\eqref{eq:anaCz} would therefore be read as differential angular readouts.
Note that the equations hold for both absolute and differential angles depending on this interpretation.

Note further that we also discarded the second-order terms ($\varphi_\text{SC}^2,\,\eta_\text{SC}^2$) in this representation. 
These become small for small \gls{S/C} jitter amplitudes. 
However, we kept the terms $\varphi_\text{SC}\varphi_i$, $\eta_\text{SC}\eta_i$, with $i\in\{1,2\}$, which are basically also second-order terms but linearised for static test mass alignment angles. 
Since the latter can significantly exceed the level of \gls{S/C} jitter (compare e.g.\ Tab.~\ref{tab:IfoCADangles}), these terms are not negligible. 

We will use the model Eq.~\eqref{eq:anamodel} in \cite{LPFdata22} for the analysis of the \gls{LPF} data.

\section{Applicability of the Results to the Mission}
\label{sec:discussion}
The direct reduction of the \gls{TTL} coupling noise by the realignment of critical components is a powerful \gls{TTL} coupling noise suppression strategy.
In \gls{LPF}, the alignment of the test masses largely affected the \gls{TTL} noise level.
Based on a simple analytical \gls{TTL} coupling model, the nominal test mass set-points have been changed three times during the mission for \gls{TTL} noise suppression. 
However, it was only possible to reduce but not fully mitigate the cross-talk making an subtraction of the residual noise necessary.

We think that the realignment would have been more successful with the analytical \gls{TTL} model that we present in this paper. 
In this section, we discuss the usability of the new model and compare it to the models used during the operation time of \gls{LPF}. 

\subsection{Comparison with Tilt-To-Length Models Used During the LISA Pathfinder Mission}
\label{sec:discussion_oldmodel}
The first TTL models for \gls{LPF} relied mainly on a simplified geometric piston model for test mass rotations and gained more complexity during the mission. However, these models did not sufficiently describe the TTL noise in \gls{LPF}. 
The new analytical model presented in this paper accounts for additional \gls{TTL} coupling mechanisms and all optical setup parameters. 
Besides the complexity, we find three significant differences between this new model and the models available during the LPF mission.

First, when interpreting the \gls{S/C} jitter as test mass jitter, we have to model the corresponding coupling for rotations about the \gls{S/C}'s centre of mass instead of the test mass centres. 
During the mission, the relevance of the location of the centre of rotation was underestimated and the models considered common-mode rotations of the test masses about their centre of mass instead.
The derivation of the new model presented in this paper showed that this simplification was erroneous.
Mainly the large longitudinal offset of the respective centre of rotation adds additional coupling.

Second, lateral shifts of the test masses were found not to change the distance between the \gls{S/C} centre of rotation and the point of reflection at the test masses. 
Consequently, the model presented here describes that lateral test mass displacements do not change the level of \gls{TTL} noise, unlike the original models.

Third, non-geometric coupling effects have often been ignored in previous models. 
However, as we show in Sec.~\ref{sec:TTLmodel_Mechanisms}, they significantly contribute to the \gls{TTL} coupling noise.

\subsection{Usability of the New Analytical Model}

Since the \gls{LPF} mission ended in 2017, which was before this model was derived, there was no direct application of the analytical model (Eq.~\eqref{eq:anamodel}) to the mission. 
However, the equations help understand the \gls{TTL} coupling mechanisms and their importance for \gls{LPF} better.
By this, we could learn why the initial attempts of deriving the realignment angles showed inconsistencies with the resulting noise changes, while later attempts were reducing the noise, but could not fully suppress it. 
Further information and particularly a detailed \gls{TTL} coupling data analysis for \gls{LPF} are given in \cite{LPFdata22}.

\section{Conclusions}
\label{sec:summary}
Within this work, we have presented an updated and most complete analytical \gls{TTL} coupling model for \gls{LPF}. For this, we considered geometric as well as non-geometric \gls{TTL} coupling effects.
The resulting model mainly differs from the models available during the mission in the assumed centre of rotation and the previously neglected non-geometric terms. An important consequence of the change of the centre of rotation is that the updated \gls{TTL} coupling models are independent from lateral realignments of the test masses. 
We could show that the updated model coincides with numerical simulations with considerable accuracy.
While we assumed exact setup parameters in the modelling and simulation, they are subject to measurement uncertainties or long-term drifts in experiment.
Therefore, we characterised how these uncertainties affect the stability of the derived coupling coefficients.
This analysis yielded relative errors of the coupling terms up to 10\%.
In the last step, we translated the length signal formulation of the \gls{TTL} coupling into an equation showing the \gls{TTL} coupling contribution to the $\Delta g$-measurements in \gls{LPF}.

From the new analytical model, we can deduce three main conclusions.
First, our investigations have shown that the lever arm and the piston effect (geometric and non-geometric contributions) were the main noise contributors to the overall \gls{TTL} coupling.
Both added angular coupling noise of the same magnitude.
Hence, a reasonable mitigation of the angular \gls{TTL} noise could only be achieved when both effects counteract each other. 

Second, the \gls{TTL} coupling model consists of first- and second-order terms. 
While a linear coupling model would, in general, not be sufficient for the characterisation of this noise, it can, for jitter of small amplitude, be linearised in the \gls{S/C} jitter parameters. 
The resulting coupling coefficients depend on both the setup and beam parameters as well as the test mass alignments.
If the case that second-order coupling becomes non-negligible due to large jitter amplitudes, we have seen that only the angular jitter yields second-order noise terms.

Last, we have shown that the geometric and non-geometric coupling contributions both add significant \gls{TTL} coupling noise to the full signal.
Therefore, a purely geometric modelling of the \gls{TTL} noise would not be sufficient, but both have to be considered in the \gls{TTL} noise analysis. 

The presented analytical model is an important basis for the investigation of the \gls{TTL} coupling measured in \gls{LPF} and the corresponding lessons learned for \gls{LISA}.
Thus, the model Eq.~\ref{eq:anamodel} will be applied to the \gls{LPF} data in \cite{LPFdata22} and \cite{LPFlongterm22}. 
We show in \cite{LPFdata22} that it can be used to explain the \gls{TTL} coupling behaviour in experiment. This analysis holds as a further validation of the presented model.
In addition, it can be used to analyse the stability of the optical setup itself. As discussed in \cite{LPFlongterm22}, our analytical \gls{TTL} coupling description can relate the long-term and temperature-depended drifts of the coupling coefficients to unconsidered test mass realignments and deformations of the \gls{LPF} optical bench.

\section*{Acknowledgements}
We thank Valerio Ferroni and Nikolaos Karnesis for valuable discussions.
This work was made possible by funds of both the German Space Agency, DLR, and the Deutsche Forschungsgemeinschaft (DFG). 
We gratefully acknowledge the German Space Agency, DLR and support by the Federal Ministry for Economic Affairs and Climate Action based on a resolution of the German Bundestag (FKZ 50OQ0501, FKZ 50OQ1601, and FKZ 50OQ1801).
Likewise, we gratefully acknowledge the Deutsche Forschungsgemeinschaft (DFG) for funding the Cluster of Excellence QuantumFrontiers (EXC 2123, Project ID 390837967).

\appendix

\section{Parameter list}
\label{sec:parameters}

The \gls{LPF} setup and beam parameters substituted into the analytical \gls{TTL} model presented in Sec.~\ref{sec:TTLmodel_LPSmodel} are summarised in Tabs.~\ref{tab:parameters_setup}, \ref{tab:parameters_angles} and \ref{tab:parameters_beam}. 
The beam parameters have previously been published in \cite{Killow2016}. The \gls{S/C} centre of mass has been defined in \cite{Armano2019_stability}.
The other parameters have been extracted from a three-dimensional \gls{LPF} model implemented in IfoCAD, which is based on \cite{TNoptocad}.
The numbers are given for the case of non-rotated test masses.
We further compute all numbers for the case of interference at the A-diode of the x12-interferometer (PD12A, see Fig.~\ref{fig:LPFifo_x12}). 

\begin{table}
  \caption{Setup parameters. 
  The distances between the beam's point of reflection (PoR) at the TM and the respective centre of rotation are given here in the LPF coordinate system to avoid confusion. Mind that for computation as introduced in \cite{G21,NG21}, the signs of the longitudinal and lateral offset for TM2 would have to be inverted: The coordinate system is set there by the propagation direction of the incident beam, i.e., opposite to the LPF coordinate system.
  The beam offsets are defined as the offset of the point where the respective beam axis hits the surface of the A-diode of the x12-interferometer from the centre of that diode's surface.}
  \label{tab:parameters_setup}
  \begin{tabular}{p{0.4\textwidth}@{\hskip 6pt}r}
  \toprule
  setup parameter & value \\%& uncertainty \\
  \midrule
  lever arm length between both TMs neglecting the windows      & 0.356\,m \\
  lever arm length between TM2 and the PD neglecting the window & 0.143\,m \\
  \midrule
  longitudinal distance between the PoR and centre of TM1 & -0.023\,m  \\
  lateral distance between the PoR and centre of TM1      & -6.37\,$\upmu$m \\
  vertical distance between the PoR and centre of TM1     &  3.42\,$\upmu$m \\
  longitudinal distance between the PoR and centre of TM2 & 0.023\,m \\
  lateral distance between the PoR and centre of TM2      & -15.8\,$\upmu$m \\
  vertical distance between the PoR and centre of TM2     & 10.5\,$\upmu$m \\
  \midrule
  longitudinal distance between the PoR at TM1 and the S/C centre of mass & 0.160\,m \\% & $-$7.3/+5.5\,mm \\
  lateral distance between the PoR at TM1 and the S/C centre of mass     &  -0.006\,m \\% & $-$5.5/+8.8\,mm \\
  vertical distance between the PoR at TM1 and the S/C centre of mass    & 0.063\,m \\% & $\pm$5\,mm \\
  longitudinal distance between the PoR at TM2 and the S/C centre of mass & -0.170\,m \\% &  $-$7.3/+5.5\,mm \\
  lateral distance between the PoR at TM2 and the S/C centre of mass     & -0.006\,m \\% & $-$5.5/+8.8\,mm \\
  vertical distance between the PoR at TM2 and the S/C centre of mass    & 0.063\,m \\% & $\pm$5\,mm \\
  \midrule
  thickness of the windows        & 6.05\,mm \\
  refraction index of the windows & 1.61     \\
  \midrule
  slit width of the PD & 45\,$\upmu$m \\
  horizontal measurement beam offset from the PD centre & -1.34\,$\upmu$m  \\
  vertical measurement beam offset from the PD centre   & 13.3\,$\upmu$m \\
  horizontal reference beam offset from the PD centre   & -2.12\,$\upmu$m \\
  vertical reference beam offset from the PD centre     & -2.26\,$\upmu$m \\
  \bottomrule
  \end{tabular}
\end{table}

\begin{table}
  \caption{Beam alignment angles. 
  The angular measurement beam alignment before hitting the TMs is provided by the propagation angles in the respective plane.
  In the case of the beam alignment angles with respect to the detector, we assume the A-diode of the x12-interferometer.}
  \label{tab:parameters_angles}
  \begin{tabular}{p{0.37\textwidth}r}
  \toprule
  beam angle & value \\
  \midrule
  tilt of incoming beam at TM1: $xy$-plane & -0.0785\,rad \\
  tilt of incoming beam at TM2: $xy$-plane &  0.0786\,rad \\
  tilt of incoming beam at both TMs: $xz$-plane & -20\,$\upmu$rad \\
  \midrule
  yaw angle between the surface normal of the PD an the negative measurement beam direction (interpreted as PD tilt)    & -165\,$\upmu$rad \\
  pitch angle between the surface normal of the PD an the negative measurement beam direction (interpreted as PD tilt)  & -20\,$\upmu$rad \\
  yaw tilt of the reference beam at the PD   & 125\,$\upmu$rad \\
  pitch tilt of the reference beam at the PD & 7\,$\upmu$rad \\
  \bottomrule
  \end{tabular}
\end{table}

\begin{table}
  \caption{The beam parameters in LPF. 
  The distances from waist are given at the point where the beam axis hits the surface of the A-diode of the x12-interferometer.}
  \label{tab:parameters_beam}
  \begin{tabular}{lr}
  \toprule
  beam parameter & value \\
  \midrule
  wave number of the beams & 5.81$\cdot10^{6}$m$^{-1}$ \\
  Rayleigh range of the measurement beam & 0.867\,m \\
  Rayleigh range of the reference beam   & 0.738\,m \\
  distance from waist of the measurement beam & 0.519\,m \\
  distance from waist of the reference beam   & 0.594\,m \\
  \bottomrule
  \end{tabular}
\end{table}
\FloatBarrier
\bibliographystyle{unsrt}
\def\bibsection{\section*{References}} 
\bibliography{References.bib}
\end{document}